# CLUSTERING BY QUANTUM ANNEALING ON THE THREE–LEVEL QUANTUM ELEMENTS QUTRITS

## V. E. Zobov and I. S. Pichkovskiy[1]

**Abstract** Clustering is grouping of data by the proximity of some properties. We report on the possibility of increasing the efficiency of clustering of points in a plane using artificial quantum neural networks after the replacement of the two-level neurons called qubits represented by the spins $S = 1/2$ by the three-level neurons called qutrits represented by the spins $S = 1$. The problem has been solved by the slow adiabatic change of the Hamiltonian in time. The methods for controlling a qutrit system using projection operators have been developed and the numerical simulation has been performed. The Hamiltonians for two well-known clustering methods, one-hot encoding and k-means ++, have been built. The first method has been used to partition a set of six points into three or two clusters and the second method, to partition a set of nine points into three clusters and seven points into four clusters. The simulation has shown that the clustering problem can be effectively solved on qutrits represented by the spins $S = 1$. The advantages of clustering on qutrits over that on qubits have been demonstrated. In particular, the number of qutrits required to represent $N$ data points is smaller than the number of qubits by a factor of $\log_2 N / \log_3 N$. Since, for qutrits, it is easier to partition the data points into three clusters rather than two ones, the approximate hierarchical procedure of data partitioning into a larger number of clusters is accelerated. At the exact data partition into more than three clusters, it has been proposed to number the clusters by the numbers of states of the corresponding multi-spin subsystems, instead of using the numbers of individual spins. This reduces even more the number of qutrits ( $N \log_3 K$ instead of $NK$ ) required to implement the algorithm.

[1] Kirensky Institute of Physics, Federal Research Center "Krasnoyarsk Science Center", Siberian Branch, Russian Academy of Sciences

# 1 Introduction

Recently, quantum mechanics has been widely used in the development of information technology, in particular, building artificial intelligence [1, 2]. The problems of optimization, pattern recognition, and data clustering can be solved more efficiently with the correct use of the quantum properties of nanoobjects. In this work, we consider the clustering problem [3–11]. Clustering is grouping the data by the proximity in the space some properties. As an example of the data, we consider points with the coordinates $(x, y)$, which represent two properties of the data, in a two-dimensional (2D) plane. The proximity of points with numbers $i$ and $j$ is characterized by the Euclidean distance $R_{ij}$ between them

$$R_{ij} = \sqrt{(x_i - x_j)^2 + (y_i - y_j)^2}, \qquad (1)$$

where $x_i$, $x_j$, $y_i$, and $y_j$ are the coordinates of points $i$ and $j$ in the Cartesian plane. The solution of the clustering problem consists in finding such a partition of a set of $N$ points into K clusters $C_\alpha$ that would minimize the value of the cost function

$$W = \frac{1}{2} \sum_{\alpha=1}^{K} \sum_{i,j \in C_\alpha} R_{ij}. \qquad (2)$$

To find the global minimum of function (2), it is necessary to enumerate a huge number of partition variants. Therefore, the clustering problem belongs to the class of NP-hard problems, which are very difficult to solve on a classical computer. Recent studies [5–11] showed that the quantum computer can solve such problems at a higher speed than a classical one. A promising quantum clustering method is quantum annealing [5–12], which is characterized by the noise immunity. This method was implemented on a D-Wave annealer [7, 10, 11]. The main problem of the available quantum computers is the limitation imposed on the number of quantum elements (qubits). The studies on reducing the number of quantum elements required to implement quantum algorithms have been carried out [5–11, 13–15]. In particular, the number of quantum elements can be reduced by passing from qubits to qudits, i.e., the $d$-level quantum elements [8, 16–27]. At the same size of computational basis $V$, the number of quantum elements decreases by a factor of $\log_2 V / \log_d V$. Unfortunately, the multilevel quantum systems become increasingly difficult to control with increasing number of levels.

In this study, we discuss the implementation of clustering on a system of three-level quantum elements (qutrits), which are represented by the spins $S = 1$. As a computational basis, we use the basis $|m_1, m_2, ..., m_n\rangle$ of the eigenfunctions of the operator $S_i^z$ of spin projections onto the $z$ axis. Each of the projections $m_i$ can take one of the three values: 1, 0, or –1. We solve the clustering problem via the slow (adiabatic) evolution of a system with the effective time-dependent Hamiltonian $0 \leq t \leq T$

$$H(t) = \left(1 - \frac{t}{T}\right) H_0 + \frac{t}{T} H_f, \qquad (3)$$

where $H_0 = h \sum_{i=1}^{n} S_i^x$ is the initial Hamiltonian of the interaction with the transverse magnetic field directed along the $x$ axis, the ground state of which can easily be prepared, and $H_f$ is the final Hamiltonian, the ground state of which encodes the solution to our problem (the minimum of function (2)). The quantum clustering algorithm has been already implemented on qubits, but has not been used on qutrits yet. In Section 2, we derive the effective time-dependent Hamiltonians for qutrits. We consider two well-known clustering methods [7]: one-hot encoding and k-means. In Section 3, we numerically simulate simple examples of clustering on qutrits.

## 2. Quantum Clustering Theory

### 2.1 One-hot encoding

In this clustering method [7, 8, 10, 11], it is necessary to search for the global minimum of function (2). As the simplest example of clustering into three clusters, we write the final Hamiltonian in the form

$$H_f = \frac{1}{2} \sum_{i,j}^{N} H_{fij}, \qquad (4)$$

$$H_{fij} = R_{ij} \left( 2 \left[ |1,1\rangle\langle 1,1|_{i,j} + |0,0\rangle\langle 0,0|_{i,j} + |-1,-1\rangle\langle -1,-1|_{i,j} \right] - 1 \right), \qquad (5)$$

where $|m_i, m_j\rangle\langle m_i, m_j|_{ij} = |m_i\rangle\langle m_i|_i \otimes |m_j\rangle\langle m_j|_j = |m_i\rangle\langle m_i|_i |m_j\rangle\langle m_j|_j$ is the projector onto the eigenfunctions of two spins at points $i$ and $j$ with the projections $m_i$ and $m_j$, respectively. We express the projection operators through the spin operators of an individual spin $i$ ($j$)

$$|-1\rangle\langle -1|_i = -S_i^z \frac{1 - S_i^z}{2}, \quad |0\rangle\langle 0|_i = 1 - (S_i^z)^2, \quad |1\rangle\langle 1|_i = S_i^z \frac{1 + S_i^z}{2}. \qquad (6)$$

As a result of the adiabatic evolution with Hamiltonian (3), the system will pass to the ground state of Hamiltonian (4), which corresponds to the minimum of function (2). The energy minimum determines the projections of the spins $m_i$ at the points. The points with the same spin projections belong to the same cluster. Since the three projections 1, 0, and −1 in Eq. (4) are equivalent, the ground state is sixfold degenerate. To remove the threefold degeneracy, one could fix a value of projection 1 at the first spin, for example, by applying a strong magnetic field to it. Thus, the calculation can be simplified via reducing the Hilbert space dimension by a factor of 3 and taking Hamiltonian (4) in the form

$$H_f = \frac{1}{2} \sum_{i,j \neq 1}^{N-1} H_{fij} + \sum_{j}^{N-1} R_{1j} \left( 2|1\rangle\langle 1|_j - 1 \right). \qquad (7)$$

Let us consider the more complex case $K > 3$. Here, additional spins should be introduced. In [7, 8, 10], each data point is associated with $K$ different spins the numbers of which number the clusters. At such an approach, $NK$ spins are required. We propose to number the clusters by the numbers of states of the multi-spin subsystems representing data points. For example, a system of two spins $S = 1$ has nine states:

$$|\psi_1\rangle = |1,1\rangle, \ |\psi_2\rangle = |1,0\rangle, \ |\psi_3\rangle = |1,-1\rangle, \ |\psi_4\rangle = |0,1\rangle, \ |\psi_5\rangle = |0,0\rangle, ..., |\psi_9\rangle = |-1,-1\rangle. \quad (8)$$

This approach follows from a method for amplitude coding of the states proposed in [5]. At such numbering, Hamiltonian (5) takes the form

$$H_{fij} = R_{ij} \left[ 2 \sum_{q=1}^{K} |\psi_q\rangle\langle\psi_q|_i \otimes |\psi_q\rangle\langle\psi_q|_j - 1 \right], \quad (9)$$

where $|\psi_q\rangle$ are the functions from sets (8) for points $i$ and $j$ at $K = 9$. In general, at $K = 3^n$, these will be the functions of a similar set for the $n$-spin system. In the case $3^{n-1} < K < 3^n$, the extra states can be removed using, e.g., the penalty Hamiltonian

$$H_p = \sum_{i}^{N-1} a_i \sum_{q=K+1}^{3^n} |\psi_q\rangle\langle\psi_q|_i, \quad (10)$$

where $a_i$ is a sufficiently large constant. The penalty Hamiltonian is the knowingly higher energy of the multi-spin states with at least one spin in the forbidden state as compared with the energy of the ground state of Hamiltonian (9). To satisfy this condition, we can take $a_i > \max R_{ij}$. There are also other variants, which are considered below on certain examples.

## 2.2 k-means++

In this method [4–7], a local minimum of function (2) is sought for. At the first stage, for each of the $K$ clusters, one point (centroid) is chosen. At the second stage, the minimum of cost function (2) is sought by enumerating all the variants of partition of the remaining $N - K$ points into clusters. The number of such variants is significantly smaller than when searching for the global minimum by the above-described method.

To eliminate the degeneracy and reduce the Hilbert space dimension, we assign fixed spin states to the chosen centroids

$$|\varphi_1\rangle, |\varphi_2\rangle, \ldots |\varphi_c\rangle, \ldots |\varphi_K\rangle. \quad (11)$$

Then, Hamiltonian (4) takes the form

$$H_f = \sum_{c=1}^{K} \sum_{j}^{N-K} R_{cj} \left( 2|\varphi_c\rangle\langle\varphi_c|_j - 1 \right), \quad (12)$$

where the first summation is made over $K$ centroids with the states assigned from list (11) and the second summation is made over the remaining points. For example, at $K = 3$, we assign the state $|1\rangle_1$ to the first centroid, the state $|0\rangle_2$ to the second centroid, and the state $|-1\rangle_3$ to the third centroid. At $K = 9$, we take the states from set (8) for the centroids. As in the above method, at $3^{n-1} < K < 3^n$, to eliminate the extra states, we should add the penalty Hamiltonian

$$H_p = \sum_j^{N-K} b_j \sum_{q=K+1}^{3^n} |\varphi_q\rangle\langle\varphi_q|_j, \qquad (13)$$

where $b_j$ is a sufficiently large constant.

### 3 Numerical Simulation

In this section, we perform the numerical simulation of the proposed quantum algorithms using simple examples. To simulate the evolution of a system with time-dependent Hamiltonian (3), we divide the total time $T$ into $M$ short intervals $\Delta t = 0.1$ and, in each of them, ignore the change in Hamiltonian (3) [27]. The solution to our problem $|\Psi\rangle$ will be sought as a product

$$|\Psi(t=T)\rangle \cong \prod_{l=0}^{M} \exp\left\{-i\Delta t \left(\frac{l}{M} H_f + \left(1 - \frac{l}{M}\right) H_0\right)\right\} |\Psi(t=0)\rangle. \qquad (14)$$

The simulation was performed at $M = 2000$.

#### 3.1 Partition of a set of six points into three clusters by the one-hot encoding method

Using a random data generator in the range from $-10$ to $10$, we obtained the coordinates of six points

$$(4, -2), (-7, 7), (6, -9), (-6, 8), (-2, -6), (-9, 5), \qquad (15)$$

and assigned numbers from 1 to 6 to them. Having calculated distances $R_{ij}$ (1) in Hamiltonian (7), we substituted them into (14) and made the calculation. The solution of this problem is shown in Fig. 1. It can be seen that our algorithm managed to solve the problem.

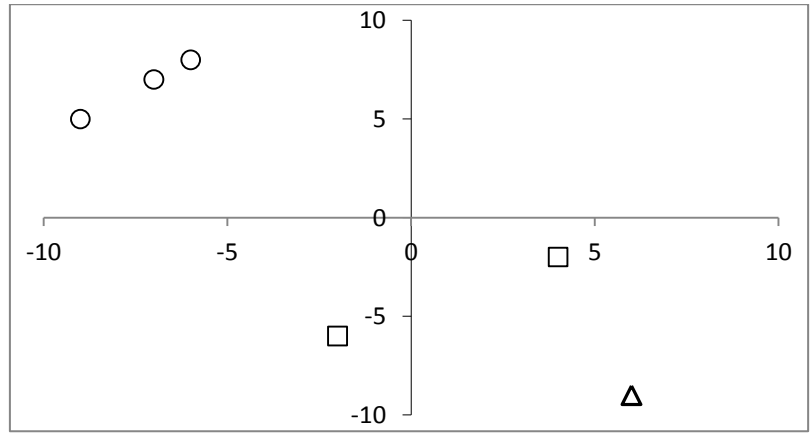

Fig. 1. Result of partition of a set of six data points into three clusters using the one-hot encoding method at $h = 2$. The first cluster: (–9, 5), (–7, 7), and (–6, 8) is shown by circles; the second cluster: (–2, –6) and (4, –2) by squares; and the third cluster (6, –9) by a triangle.

**3.2 Partition of a set of six points into two clusters by the one-hot encoding method**

As in the previous case, using a random number generator, we obtained the coordinates of six points

$$(6, 6), (-6, 5), (-3, 9), (4, -10), (-7, 4), (-5, 1). \quad (16)$$

The simulation is performed similarly, but (5) is added with the penalty Hamiltonian.

$$H_{fij} = 2R_{ij}\left(\left|-1\right\rangle\left\langle-1\right|_i + \left|-1\right\rangle\left\langle-1\right|_j\right), \quad (17)$$

which imposes a penalty on all the states with the spin projection $S^Z = -1$, specifically, $\left|1,-1\right\rangle$, $\left|0,-1\right\rangle$, $\left|-1,1\right\rangle$, and $\left|-1,0\right\rangle$. The result is shown in Fig. 2.

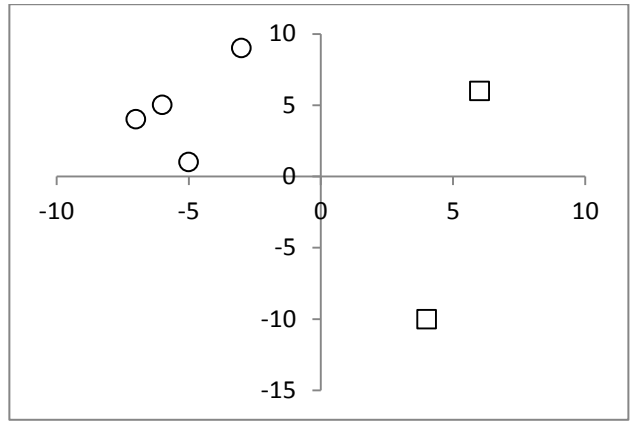

Fig. 2. Result of partition of a set of six data points into two clusters using the one-hot encoding method at $h = 8$. The first cluster (–7, 4), (–6, 5), (–5, –1), and (–3, 9) is shown by circles and the second cluster (4, –10) and (6, 6) by squares;

**3.3 Partition of a set of nine points into three clusters by the k-means++ method**

Among the nine points with the random coordinates

$$(8, -1), (-2, -6), (1, 6), (4, -4), (3, 8), (9, -4), (-5, 8), (-6, -8), (3, -10), \quad (18)$$

we will use the points (8, –1), (–2, –6), and (1, 6) as centroids. We assign them numbers 1, 2, and 3 and the states $\left|1\right\rangle_1$, $\left|0\right\rangle_2$, and $\left|-1\right\rangle_3$ respectively. The six remaining points (4, –4), (3, 8), (9, –4), (–5, 8), (–6, –8), and (3, –10) numbered sequentially will be distributed over these clusters. The final Hamiltonian for this problem takes the form

$$H_f = \sum_{j=1}^{6} R_{1j}\left(2\left|1\right\rangle\left\langle1\right|_j - 1\right) + \sum_{j=1}^{6} R_{2j}\left(2\left|0\right\rangle\left\langle0\right|_j - 1\right) + \sum_{j=1}^{6} R_{3j}\left(2\left|-1\right\rangle\left\langle-1\right|_j - 1\right). \quad (19)$$

Substituting this Hamiltonian into Eq. (14), we find the solution to the clustering problem shown in Fig. 3. It can be seen that our algorithm managed to solve the problem.

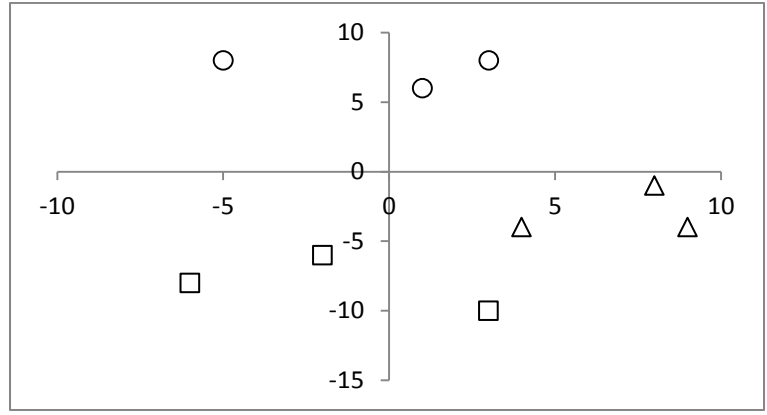

Fig. 3. Result of clustering of a set of nine data points into three clusters using the k-means++ method at $h = 8$. The first cluster: (–5, 8), (1, 6), and (3, 8) is shown by circles; the second cluster: (–6, –8), (–2, –6), and (3, –10) by squares; and the third cluster: (4, –4), (8, –1), and (9, –4), by triangles.

### 3.4 Partition of a set of seven points into four clusters by the k++ method

We take 7 point with the random coordinates

$$(-9, 10), (1, 9), (-8, -3), (-2, -9), (4, -2), (8, -8), (10, -5), \tag{20}$$

Since the partition into four clusters is needed, each point $i$ is encoded in the four states of two spins

$$|\varphi_1\rangle_i = |1,1\rangle_i, \ |\varphi_2\rangle_i = |1,0\rangle_i, \ |\varphi_3\rangle_i = |1,-1\rangle_i, \ |\varphi_4\rangle_i = |0,1\rangle_i. \tag{21}$$

We remove the remaining five states of the two-spin systems using penalty Hamiltonian (13).

As centroids, we choose four points (–9, 10), (1, 9), (–8, –3), and (4, –2) and assign them the sequential numbers and states (21), according to the numbers. The remaining three points (–2, –9), (8, –8), and (10, –5) are numbered sequentially. To group them, we obtain the Hamiltonian

$$\begin{aligned}H_f = &\sum_{j=1}^{3} R_{1j}\left(2|1,1\rangle\langle 1,1|_j - 1\right) + \sum_{j=1}^{3} R_{2j}\left(2|1,0\rangle\langle 1,0|_j - 1\right) + \\ &+ \sum_{j=1}^{3} R_{3j}\left(2|1,-1\rangle\langle 1,-1|_j - 1\right) + \sum_{j=1}^{3} R_{4j}\left(2|0,1\rangle\langle 0,1|_j - 1\right)\end{aligned}, \tag{22}$$

where the distance is calculated using Eq. (1). Substituting Hamiltonian (22), together with penalty Hamiltonian (13) rewritten in the corresponding functions, into Eq. (14), we find the solution to the clustering problem (Fig. 4).

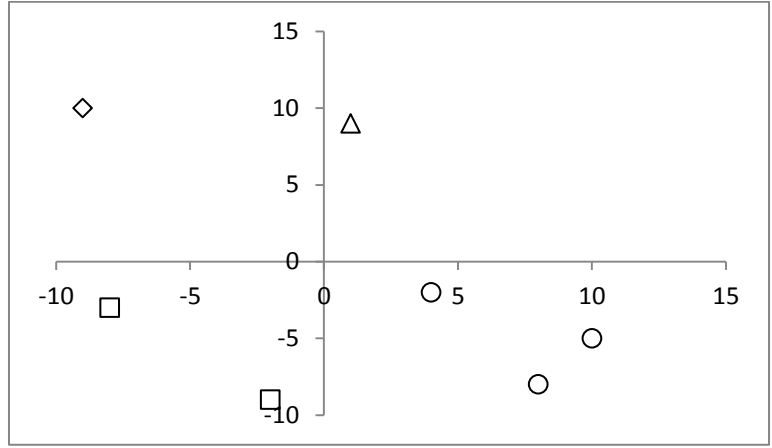

Fig. 4. Result of clustering of a set of seven data points into four clusters using the k-means++ method at $h=8$. The first cluster: (–9, 10) is shown by a rhomb; the second cluster: (–8, –3) and (–2, –9), by squares; the third cluster: (1, 9) by a triangle; and the forth clusters (4, –2), (8, –8), (10, –5), by circles.

## 4. Discussion

Controlling a system of the spins $S=1$ (qutrits) differs from controlling a system of the spins $S=1/2$ (qubits), first of all, due to the state with the projection $S_z=0$, which is not affected by a magnetic field. Therefore, we used the technique of projectors in the Hamiltonians. In this representation, we derived the formulas required for controlling a system of qutrits during implementation of the adiabatic quantum clustering algorithm. We grouped data points in a plane using two methods: (i) the one-hot encoding method, in which all possible variants of partitioning a set of points into a specified number of clusters are enumerated and (ii) the k-means++ method, in which, first, one point (centroid) is selected for each cluster and, then, the remaining points are distributed between the clusters. Our calculation showed that the clustering problem can be solved on qutrits represented by the spins $S=1$. Moreover, clustering on qutrits has advantages over that on qubits. First, the number of qutrits required to represent $N$ data points is smaller than the number of qubits by a factor of $\log_2 N / \log_3 N$. Second, for qubits, the simplest thing is to partition data points into two clusters, while for qutrits, into three clusters. Therefore, the approximate hierarchical procedure of data partition into a larger number of clusters is accelerated. As was shown above, at the exact partition of data into more than three clusters, the number of qutrits required for implementing the algorithm can be further reduced ($\log_2 N / \log_3 N$ instead of $NK$) if we refuse the conventional cluster numbering procedure [7, 8, 10]. We proposed to number the clusters by the numbers of states of the corresponding multi-spin systems instead of the numbers of individual spins. Note that this numbering procedure can also be applied to qubits.

Various three-level quantum systems can be used as qutrits, including photons [25], atoms and ions in traps [17, 25], superconducting systems [25, 26], and objects with the spin $S = 1$ in magnetic and crystal fields. The latter include quadrupole nuclei [16, 19, 20] of deuterium, nitrogen, or lithium, as well as the NV centers in diamond (paramagnetic color centers formed by electrons on vacancies near nitrogen atoms) [23, 24, 28]. The variant with the NV centers is preferred due to the presence of a strong dipole–dipole interac-tion between them, which is necessary to implement conditional operations in quantum algorithms. In addition, in this case, the frequencies of transitions between different energy levels are strongly different, which makes it possible to control the states of a system using the transition-selective rf field pulses. The Hamiltonian in a form suitable for the experimental implementation is obtained by substituting the products of projectors expressed in spin operators (6) into $H_f$ and $H_p$. It should be noted that, along with the two-spin interaction in the form of the Ising or dipole–dipole interaction, the multi-spin interactions and the interactions containing squared spin operators will be required. The methods for obtaining such interactions using the rotation operators selective with respect to the transitions between levels of each of the spins were described in [21–24].

## Acknowledgments

This study was supported by the Theoretical Physics and Mathematics Advancement Foundation "BASIS". We are grateful for their trust and assistance in research.